\documentclass{ws-ijmpcs}

\begin{document}

\markboth{Ju-Jun Xie, A. Mart\'inez Torres and E. Oset}
{$N^*(1920)(1/2^+)$ state in the $NK\bar{K}$ system}

%
\catchline{}{}{}{}{}
%

\title{$N^*(1920)(1/2^+)$ state in the $NK\bar{K}$ system}

\author{Ju-Jun Xie}

\address{Institute of modern physics, Chinese Academy of Sciences, Lanzhou 730000, China \\
State Key Laboratory of Theoretical Physics, Institute of
Theoretical Physics, Chinese Academy of Sciences, Beijing 100190, China \\
xiejujun@impcas.ac.cn}

\author{A. Mart\'inez Torres}

\address{Instituto de F\'isica, Universidade de S$\tilde{a}$o Paulo, C.P. 66318, 05389-970 S$\tilde{a}$o Paulo, SP, Brazil}

\author{E. Oset}

\address{Departamento de F\'{\i}sica Te\'{o}rica and IFIC, Centro Mixto CSIC-Universidad de Valencia,
Institutos de Investigaci\'on de Paterna, Aptd. 22085, E-46071
Valencia, Spain}

\maketitle


\begin{abstract}

We study the three body $N \bar{K} K$ system by using the fixed
center approximation to the Faddeev equations, taking the
interaction between $N$ and $\bar{K}$, $N$ and $K$, and $\bar{K}$
and $K$ from the chiral unitary approach. Our results suggest that a
$N\bar{K}K$ hadron state, with spin-parity $J^P=1/2^+$, and mass
around $1920$ MeV, can be formed.

\keywords{Faddeev fixed-center approximation; $NK\bar{K}$ system;
$N^*(1920)$ resonance.}
\end{abstract}

\ccode{PACS numbers: 11.25.Hf, 123.1K}

\section{Introduction}

The study of hadron structure is one of the important issues in
contemporary hadron-nuclear physics and is attracting much
attention. For example, the $\Lambda(1405)$ state, which is
catalogued as a four-star $\Lambda$ resonance in the Particle Data
Group (PDG) review book~\cite{pdg2012}, has structure and properties
which are still controversial. Within the unitary chiral theory, two
$\Lambda(1405)$ states are dynamically
generated~\cite{osetkbarN,reciokbarN,hyodokbarN,hyodoweisekbarN,jido1405,ollerchiral1,borasoychiral1,ollerchiral2,borasoychiral2}.
The heavier one corresponds to basically a $\bar{K}N$ bound state
and the lighter one is more looking like a $\pi \Sigma$ resonance.
For mesonic resonances, the $f_0(980)$ and $a_0(980)$ are also
dynamically generated from the interaction of $\bar{K}K$, $\pi \pi$,
and $\eta \pi$ treated as coupled channels in $I=0$ and $I=1$,
respectively~\cite{ollerkbark1,ollerkbark2,ollerkbark3,nicolakbark,pelaezkbark,kaiserkbark,markushinkbark}.

For the three body $N \bar{K} K$ system, it is naturally expected
that the three hadrons $N \bar{K} K$ form a bound state because of
the strong attraction in the $\bar{K} N$ and $\bar{K} K$ subsystems.
Indeed, this state has been studied with nonrelativistic three-body
variational calculations,~\cite{jido1920} and by solving the Faddeev
equations in a coupled channel approach~\cite{alberto1920}. They all
found a bound state of the $N \bar{K}K$ system with total isospin
$I=1/2$ and spin-parity $J^P=1/2^+$.

Along this line, in the present work, we reinvestigate the
three-body $N \bar{K} K$ system by considering the interaction of
the three particles among themselves. With the two-body $N\bar{K}$,
$NK$ and $K\bar{K}$, $KN$ scattering amplitudes from the chiral
unitary approach, we solve the Faddeev equations by using the Fixed
Center Approximation (FCA), which has been used before, in
particular in the study of the $\bar{K} d$ interaction at low
energies~\cite{Chand:1962ec,Barrett:1999cw,Deloff:1999gc,Kamalov:2000iy}.
This approach was also used to describe the $f_2(1270)$,
$\rho_3(1690)$, $f_4(2050)$, $\rho_5(2350)$ and $f_6(2510)$
resonances as multi-$\rho(770)$ states~\cite{fcarhorho}, and also to
study the $K^*_2(1430)$, $K_3^*(1780)$, $K^*_4(2045)$,
$K^*_5(2380)$, and $K^*_6$ resonances as $K^*-$multi$-\rho$
states~\cite{YamagataSekihara:2010qk}. Furthermore, it has been
recently argued that the $\Delta_{5/2^+}$ should be interpreted
instead as two distinctive resonances based on a solution of the
$\pi \Delta \rho$ system by using the FCA~\cite{Xie:2011uw}.

\section{Formalism }

We consider the $f_0/a_0(980)$ scalar meson as a bound state of
$\bar{K} K$ in one case, and the $\Lambda(1405)$ state as a bound
state of $\bar{K} N$ in the other, which allows us to use the FCA to
solve the Faddeev equations. The analysis of the
$N-(\bar{K}K)_{f_0/a_0(980)}$ and $K-(\bar{K}N)_{\Lambda(1405)}$
scattering amplitudes will allow us to study dynamically generated
resonances.

For the case of the $K-(\bar{K}N)_{\Lambda(1405)}$ configuration,
the $K$ is assumed to scatter successively with the $\bar{K}$ and
$N$. Then the FCA equations are written in terms of two partition
functions $T_1$ and $T_2$, which sum up to the total three body
scattering amplitude $T_{K\Lambda(1405) \to K\Lambda(1405)}$,
\begin{eqnarray}
 T_1 = t_1+t_1G_0T_2, ~~~  T_2 = t_2+t_2G_0T_1, ~~~   T_{K\Lambda(1405) \to K\Lambda(1405)} = T_1+T_2,
 \nonumber
\end{eqnarray}
where $t_i$ represent the $K\bar{K}$ and $KN$ unitarized scattering
amplitudes, and $G_0$ is the loop function for the $K$ meson
propagating inside the $\Lambda(1405)$ cluster (see
Ref.~\cite{Xie:2010ig} for more details).

\section{Results and discussion}

\begin{figure}[htbp]
\begin{center}
\includegraphics[scale=0.31]{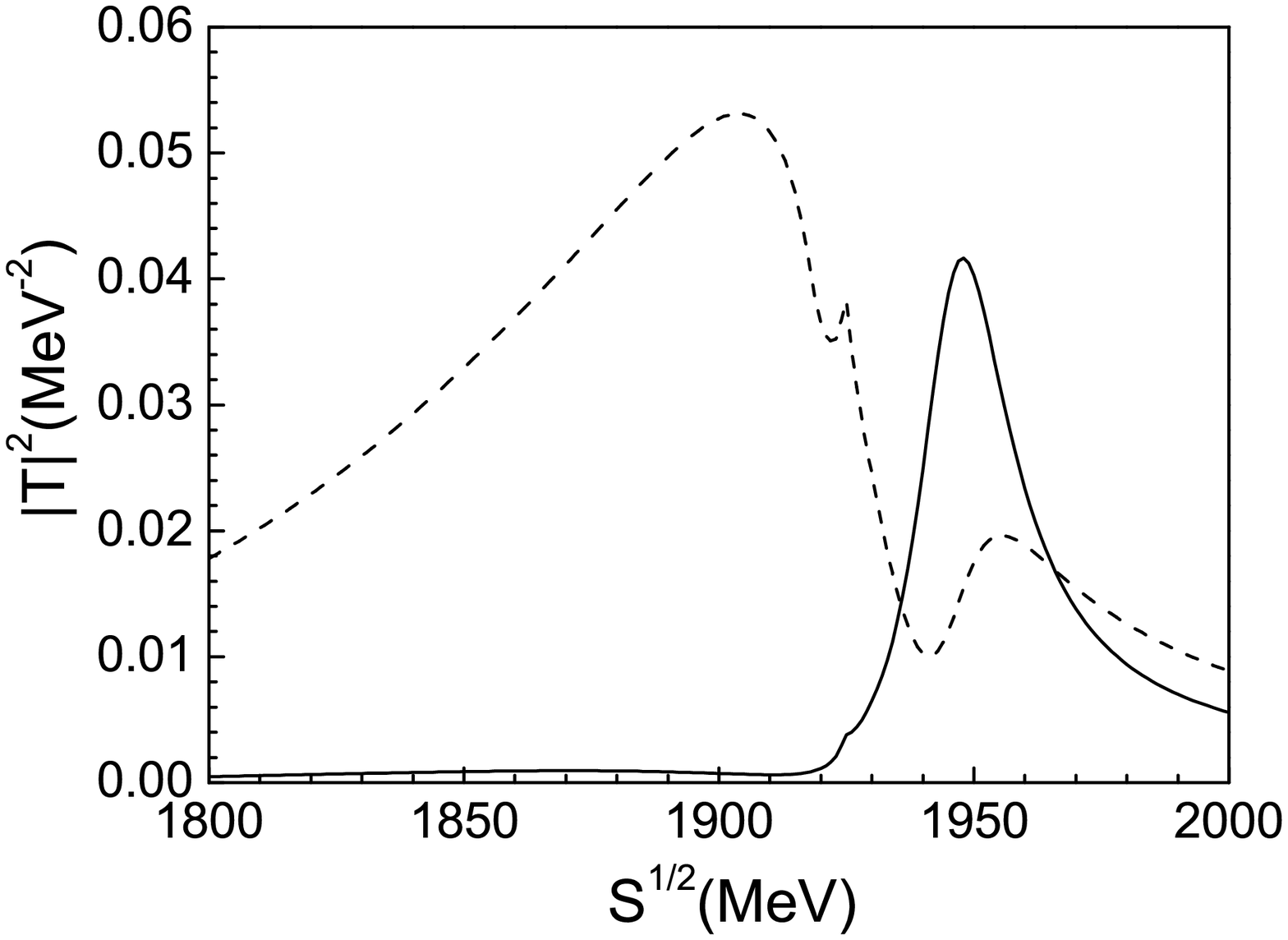}%
\includegraphics[scale=0.31]{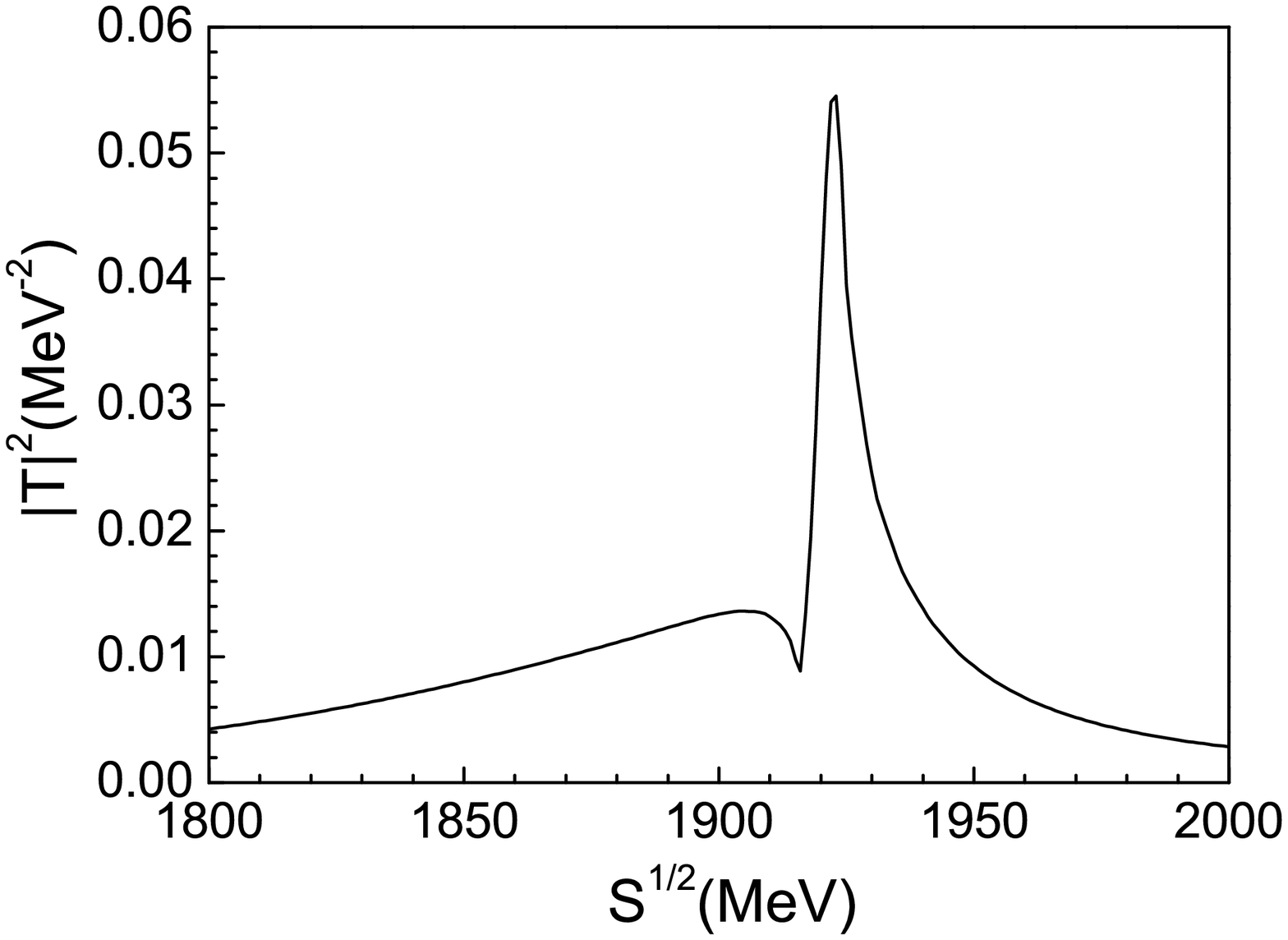}  %
\vspace{-0.5cm} \caption{Modulus squared of the $Nf_0(a_0(980))$ and
the $K\Lambda(1405)$ scattering amplitude in $I_{\text{total}}=1/2$.
Left: solid line and dashed line stand for the $Nf_0 \to Nf_0$
scattering and the $Na_0 \to Na_0$ scattering. Right:
$K\Lambda(1405) \to K\Lambda(1405)$ scattering.}
\label{Fig:1half980}%
\end{center}
\end{figure}

We calculate the three body scattering amplitude $T$ with total
isospin $I=1/2$ and spin-parity $J^P=1/2^+$ and associate the peaks
in the modulus squared $|T|^2$ to resonances. In
Fig.~\ref{Fig:1half980}, we show our results for the modulus squared
$|T|^2$. The results show a clear peak in the case of $Na_0 \to
Na_0$ around $1915$ MeV. For the $K\Lambda(1405) \to K\Lambda(1405)$
scattering, we see a clear peak around $1925$ MeV. The strength of
$|T|^2$ at the peak is similar to that for the $Na_0 \to Na_0$
scattering. With the proper comparison: $T_{Na_0 \to Na_0}$ versus
$\frac{M_{a_0(980)}}{m_K}T_{K\Lambda(1405) \to K\Lambda(1405)}$, we
obtain $|\frac{M_{a_0(980)}}{m_K} T_{K\Lambda(1405) \to
K\Lambda(1405)}|^2 \simeq 4 |T_{Na_0 \to Na_0}|^2$, which indicates
that the preferred configuration is $K\Lambda(1405)$.

From our results, the clear peak around $1920$ MeV in the scattering
amplitude for the $N\bar{K}K$ system, indicates that we have a
resonant state made of these components. Furthermore, the main
$K\Lambda(1405)$ component over the $Nf_0(a_0(980))$ serves to put
the peaks with moderate strength around $1950$ MeV seen in
Fig.~\ref{Fig:1half980} in a proper context, indicating that the
effect of this configuration in that energy region can be diluted
when other large components of the wave functions are considered,
such that we should not expect that these peaks would have much
repercussion in any physical observable.

\section{Conclusions}

We have performed a calculation for the three body $N\bar{K}K$
scattering amplitude by using the FCA to the Faddeev equations,
taking the interaction between $N$ and $\bar{K}$, $N$ and $K$, and
$\bar{K}$ and $K$ from the chiral unitary approach. It is found that
in both $Na_0(980)$ and $K\Lambda(1405)$ configurations there is a
clear peak around $1920$ MeV indicating the formation of a resonant
$N\bar{K}K$ state around this energy. This result is in agreement
with those obtained in previous
calculations~\cite{jido1920,alberto1920}, which support the
existence of a $N^*$ state with spin-parity $J^P=1/2^+$ around
$1920$ MeV. We also found that the $K\Lambda(1405)$ configuration is
the dominant one, where the $\bar{K}K$ subsystem can still couple to
the $f_0(980)$ and $a_0(980)$ resonances.

\section*{Acknowledgments}

A. Mart\'inez Torres thanks the financial support from the brazilian
funding agency FAPESP. This work is partly supported by the Spanish
Ministerio de Economia y Competitividad and European FEDER funds
under the contract number FIS2011-28853-C02-01 and
FIS2011-28853-C02-02, and the Generalitat Valenciana in the program
Prometeo, 2009/090. We acknowledge the support of the European
Community-Research Infrastructure Integrating Activity Study of
Strongly Interacting Matter (acronym HadronPhysics3, Grant Agreement
n. 283286) under the Seventh Framework Programme of EU. This work is
also partly supported by the National Natural Science Foundation of
China under grant 11105126.



\begin{thebibliography}{00}  
%
\bibitem{pdg2012} J. Beringer \emph{et al.} (Particle Data Group), Phys. Rev. \textbf{D 86}, 010001
(2012).
%
\bibitem{osetkbarN}E.~Oset and A.~Ramos, Nucl.\ Phys.\ A
\textbf{635}, 99 (1998).
%
\bibitem{reciokbarN} C.~Garcia-Recio, J.~Nieves,
E.~Ruiz Arriola and M.~J.~Vicente Vacas, Phys.\ Rev.\ D \textbf{67},
076009 (2003).
%
\bibitem{hyodokbarN}T.~Hyodo, S.~I.~Nam, D.~Jido and A.~Hosaka, Phys.\
Rev.\ C \textbf{68}, 018201 (2003).
%
%
\bibitem{hyodoweisekbarN}Tetsuo Hyodo and Wolfram Weise, Phys.\ Rev.\ C
\textbf{77}, 035204 (2008).
%
\bibitem{jido1405}D.~Jido, J.~A.~Oller, E.~Oset, A.~Ramos, and
U.~G.~Mei{\ss}ner, Nucl.\ Phys.\ A \textbf{725}, 181 (2003).
%
%
\bibitem{ollerchiral1}J.~A.~Oller and U.-G.~Mei{\ss}ner, Phys.\
Lett.\ B \textbf{500}, 263 (2001).
%
\bibitem{borasoychiral1}B. Borasoy, R. Nissler and W. Weise, Eur.\
Phys.\ J.\ A \textbf{25}, 79 (2005).
%
\bibitem{ollerchiral2}J. A. Oller, Eur.\
Phys.\ J.\ A \textbf{28}, 63 (2006).
%
\bibitem{borasoychiral2}B. Borasoy, U.-G.~Mei{\ss}ner, and R.
Nissler, Phys.\ Rev.\ C \textbf{74}, 055201 (2006).
%
\bibitem{ollerkbark1}J.~A.~Oller, and E.~Oset, Nucl.\ Phys.\ A
\textbf{620}, 438 (1997).
\bibitem{ollerkbark2}
  J.~A.~Oller, E.~Oset and J.~R.~Pelaez,
  Phys.\ Rev.\ Lett.\  {\bf 80}, 3452 (1998).
%
\bibitem{ollerkbark3}J.~A.~Oller, and E.~Oset, Phys.\ Rev.\ D
\textbf{60}, 074023 (1999).
%
\bibitem{nicolakbark}A.~G\'omez Nicola, and J.~R.~Pel\'aez, Phys.\ Rev.\ D
\textbf{65}, 054009 (2002).
\bibitem{pelaezkbark}
 J.~R.~Pelaez, G.~Rios,
 Phys.\ Rev.\ Lett.\  {\bf 97}, 242002 (2006).
%
\bibitem{kaiserkbark}N.~Kaiser, Eur.\ Phys.\ J.\ A \textbf{3}, 207
(1998).
%
\bibitem{markushinkbark}V.~E.~Markushin, Eur.\ Phys.\ J.\ A
\textbf{8}, 389 (2000).

%
\bibitem{jido1920}D.~Jido and Y.~Kanada-En'yo, Phys.\ Rev.\ C
\textbf{78}, 035203 (2008).
%
\bibitem{alberto1920}A.~Mart\'inez Torres, K.~P.~Khemchandani, U.~G.~Mei{\ss}ner, and
E.~Oset, Eur.\ Phys.\ J.\ A \textbf{41}, 361 (2009); A.~Mart\'inez
Torres, K.~P.~Khemchandani, and E.~Oset, Phys.\ Rev.\ C \textbf{79},
065207 (2009); A.~Mart\'inez Torres and D.~Jido, Phys.\ Rev.\ C
\textbf{82}, 038202 (2010).

\bibitem{Chand:1962ec}
 R.~Chand and R.~H.~Dalitz,
 Annals Phys.\  {\bf 20}, 1 (1962)

\bibitem{Barrett:1999cw}
 R.~C.~Barrett and A.~Deloff,
 Phys.\ Rev.\  C {\bf 60}, 025201 (1999).
\bibitem{Deloff:1999gc}
 A.~Deloff,
 Phys.\ Rev.\  C {\bf 61}, 024004 (2000).


\bibitem{Kamalov:2000iy}
 S.~S.~Kamalov, E.~Oset and A.~Ramos,
 Nucl.\ Phys.\  A {\bf 690}, 494 (2001).
%
\bibitem{fcarhorho}L.~Roca, and E.~Oset, Phys.\ Rev.\ D \textbf{82}, 054013 (2010).
\bibitem{YamagataSekihara:2010qk}
  J.~Yamagata-Sekihara, L.~Roca and E.~Oset,
  Phys.\ Rev.\  D {\bf 82}, 094017 (2010).

\bibitem{Xie:2011uw}
  J.~-J.~Xie, A.~Martinez Torres, E.~Oset and P.~Gonzalez,
  Phys.\ Rev.\ C {\bf 83}, 055204 (2011). 


\bibitem{Xie:2010ig}
  J.~-J.~Xie, A.~Martinez Torres and E.~Oset,
  Phys.\ Rev.\ C {\bf 83}, 065207 (2011). 


\end{thebibliography}
\end{document}